\documentstyle[12pt]{article}
\begin{document}
\def\thefootnote{\fnsymbol{footnote}}
\begin{flushright}
KANAZAWA-01-04  \\ 
May, 2001
\end{flushright}
\vspace*{2cm}
\begin{center}
{\LARGE\bf A light sterile neutrino based on the seesaw mechanism}\\
\vspace{1 cm}
{\Large  Daijiro Suematsu}
\footnote[1]{e-mail:suematsu@hep.s.kanazawa-u.ac.jp}
\vspace {0.7cm}\\
$^\ast${\it Institute for Theoretical Physics, Kanazawa University,\\
        Kanazawa 920-1192, JAPAN}
\end{center}
\vspace{2cm}
{\Large\bf Abstract}\\  
We propose a simple model of the neutrino mass matrix which can 
explain the solar and atmospheric neutrino problems in a 
3($\nu_L$+$\nu_R$) framework. Assuming that only two right-handed 
neutrinos are heavy and a Dirac mass matrix has a special texture, 
we construct a model with four light neutrinos. The favorable 
structure of flavor mixings and mass eigenvalues required by those 
neutrino deficits is realized as a result of the seesaw mechanism. 
Bi-maximal mixing structure might be obtainable in this scheme.
Since it contains a light sterile neutrino, it has a chance to 
explain the LSND result successfully. We consider an embedding of this
scenario for the neutrino mass matrix into the SU(5) grand unification 
scheme using the Froggatt-Nielsen mechanism based on 
U(1)$_{F_1}\times$U(1)$_{F_2}$. Both a small mixing angle MSW solution 
and a large mixing angle MSW solution are obtained for the solar 
neutrino problem depending on the charged lepton mass matrix. 
\newpage
\setcounter{footnote}{0}
\def\thefootnote{\arabic{footnote}}
\def\romth{I\hspace*{-0.8mm}I\hspace*{-0.8mm}I}
\def\romfo{I\hspace*{-0.8mm}V}
\def\romtw{I\hspace*{-0.7mm}I}

\noindent
{\Large\bf 1.~Introduction}

Recently the existence of non-trivial lepton mixing has been strongly
suggested through the atmospheric and solar neutrino observations 
whose results can be explained by assuming the neutrino oscillations 
\cite{oscil1,oscil2,sk}. 
The predicted flavor mixing is much bigger than the one of quark
sector. The explanation of this feature is a challenging issue for the
construction of a satisfactory grand unified theory (GUT) and a lot of 
works have been done
\cite{gut,af}. In most of them the smallness of the neutrino mass 
is explained by the celebrated seesaw mechanism \cite{seesaw}
and the flavor mixing structure is considered to be controlled by the
Froggatt-Nielsen mechanism \cite{fn}. There are many works in
which the Abelian flavor symmetry is discussed \cite{fnap}.
On the other hand, there is another experimental suggestion on the
neutrino oscillation by the Liquid Scintillator Neutrino Detector 
(LSND) \cite{lsnd}.
If we impose the simultaneous explanation of the result together with
the atmospheric and solar neutrino deficits, it has been well-known that 
three different values of the squared mass difference are necessary.
Then four light neutrinos including a sterile neutrino ($\nu_s$) are 
required \cite{lsnd1}. Various models of the sterile neutrino can be found
in refs.~[10-14].
Following the recent Super-Kamiokande analysis of the solar
neutrino, the explanation of the solar neutrino problem 
based on the $\nu_e$-$\nu_s$ oscillation seems to be disfavored \cite{sk}.
It suggests that the (3+1)-neutrino spectrum might be a more favored
scenario for the neutrino mass hierarchy than the (2+2)-scheme \cite{31sch}.

In this paper we consider a neutrino mass matrix in a
3($\nu_L$+$\nu_R$) framework by using the seesaw mechanism. 
However, being different from the ordinary seesaw models 
our model contains a light right-handed neutrino as a result of the special
texture of a right-handed Majorana neutrino mass matrix.
Although there are the similar works in this direction, in most of them 
it is necessary to introduce the Majorana masses for the left-handed neutrinos 
in order to obtain simultaneously the required values of the mass 
eigenvalues and the flavor mixing angles as it can be found, for example, 
in \cite{sterile1,sterile2}. 
It means that an introduction of a new triplet Higgs field might be necessary.
In the present model we only need the Dirac neutrino masses and the  
right-handed Majorana neutrino masses if we assume a special but simple 
texture in both of them at tree level. The model seems to have less 
parameters as compared to the previous ones.
 
One of the interesting points of the model is that
the large mixing angle MSW solution for the solar neutrino
problem can be consistently accommodated in the same way as other
solutions \cite{bimaxim,sma}. 
The LSND result might be also explained if we take an appropriate
solution for the solar neutrino problem \cite{sma}. 
Moreover, it is interesting that this scenario for the neutrino 
mass matrix could 
also be embedded into the GUT scheme by introducing a suitable flavor 
symmetry. Such an example in the SU(5) model will be constructed by 
fixing the charge assignment of quarks and leptons for that symmetry.

The organization of this paper is as follows.
In section 2 we define our model and discuss its various
phenomenological features in the case that the charged lepton mass
matrix is diagonal.
In section 3 we consider the embedding of the scenario into the SU(5)
GUT scheme. We discuss the realization of the required form of 
the mass matrix in the basis of the Froggatt-Nielsen mechanism.
The flavor structure in the quark sector is also discussed here.
Section 4 is devoted to the summary. 
\vspace*{5mm}

\noindent
{\Large\bf 2.~A model of neutrino mass matrix}

We consider a model defined by the following neutrino mass terms which
are different from the usual seesaw model in the 3($\nu_L$+$\nu_R$) framework:
\begin{equation}
-{\cal L}_{\rm mass}=\sum_\alpha\sum_{p=2,3}m_{p\alpha}N_p\nu_\alpha 
+\sum_{p=2,3}m_{p1}N_pN_1 +{1\over 2}\sum_{p=2,3}M_{p}N_pN_p +{\rm h.c.},
\label{eqa}
\end{equation}
where $\nu_\alpha$ is an active neutrino $(\alpha=e,\mu,\tau)$ and
$N_P~(P=1\sim 3)$ is a charge conjugated state of the 
right-handed neutrino.
We make the following assumption for the mass parameters 
in eq. (\ref{eqa}):
\begin{eqnarray}
&&m_{2e}=m_{2\mu }=m_{2\tau }\equiv \hat\eta,\quad
m_{3e}\equiv \bar\eta_1,\quad  
m_{3\mu }=m_{3\tau }\equiv \bar\eta_2,\nonumber \\
&&\hat\eta \sim \bar\eta_1\sim \bar\eta_2 < m_{21}\sim m_{31} \ll 
M_{2}\sim M_{3},
\label{eqb}
\end{eqnarray}
where the mass parameters should be understood as their
absolute values, although it is not expressed explicitly. 
A crucial difference from the usual seesaw model is that one of the 
right-handed neutrinos is assumed to be very light 
and also has very small mixings with
other heavy right-handed neutrinos.
We assume $M_{23}=0$ in the Majorana mass matrix of $N_P$ here, 
for simplicity. 
Following arguements are not largely changed 
even if we introduce the non-zero $M_{23}$. 
Under this assumption we can integrate out heavy right-handed neutrinos $N_p$
and get the following $4\times 4$ matrix as a result of the seesaw 
mechanism\footnote{It should be noted that the number of light sterile
neutrinos is restricted at most to one in the present scenario.
We obtain a $3\times 3$ matrix if all of $N_P$ are
heavy. Even in such a case as far as we assume a proportional relation
between $(m_{1\alpha })$ and $(m_{2\alpha })$ as vectors whose
components are labeled by $\alpha$, the texture for the active 
neutrinos is the same as eq.~(\ref{eqc}). Then it can be applied 
to the explanation of the solar
and atmospheric neutrino problems in the same way as the following 
discussion. It is essentially the same as 
the one discussed in ref.~\cite{gaugino}, although it is derived in the
different context.},
\begin{equation}
m_\nu=\left(\begin{array}{cccc}
A&B&B&D\\  B&C&C&E\\  B&C&C&E\\ D&E&E&F\\
\end{array}\right).
\label{eqc}
\end{equation}
The matrix elements $A\sim F$ are expressed by the model parameters in
(\ref{eqb}) as
\begin{eqnarray}
&&A={\hat\eta^2 \over M_2}+ {\bar\eta_1^2 \over M_3}, \quad 
B={\hat\eta^2 \over M_2}+ {\bar\eta_1\bar\eta_2 \over M_3},  \quad
C={\hat\eta^2 \over M_2}+ {\bar\eta_2^2 \over M_3}, \nonumber \\
&&D={\hat\eta m_{21} \over M_2}+ {\bar\eta_1 m_{31} \over M_3}, \quad
E={\hat\eta m_{21} \over M_2}+ {\bar\eta_2 m_{31} \over M_3}, \quad
F={m_{21}^2 \over M_2}+ { m_{31}^2 \over M_3}.
\label{eqd}
\end{eqnarray}
If we define the diagonalization matrix $U$ of the matrix (\ref{eqc}) 
as $m_\nu^{\rm diag}=U^Tm_\nu U$, we find that $U$ can be
written as 
\begin{equation}
U=\left(\begin{array}{cccc}
\cos\theta & -\sin\theta & 0 & -\sin\theta\sin\delta+\cos\theta\sin\gamma \\
{1\over\sqrt 2}\sin\theta& {1\over\sqrt 2}\cos\theta &  -{1\over\sqrt 2} &
{1\over\sqrt 2}(\cos\theta\sin\delta+\sin\theta\sin\gamma) \\
{1\over\sqrt 2}\sin\theta& {1\over\sqrt 2}\cos\theta &  {1\over\sqrt 2} &
{1\over\sqrt 2}(\cos\theta\sin\delta+\sin\theta\sin\gamma) \\
-\sin\gamma & -\sin\delta & 0 & 1 \\
\end{array}\right),
\label{eqe}
\end{equation}
where $\vert\sin\gamma\vert,~\vert\sin\delta\vert \ll 1$ is assumed and
mixing angles are defined by
\begin{equation}
\tan 2\theta ={2\sqrt 2B\over A-2C}, \quad 
\sin\gamma\simeq  {D\cos\theta+\sqrt 2E\sin\theta\over F}, \quad
\sin\delta \simeq {-D\sin\theta+\sqrt 2E\cos\theta\over F}.
\label{eqf}
\end{equation}
The mass eigenvalues of $m_\nu$ are expressed as
\begin{eqnarray}
&&m_1\simeq A\cos^2\theta+\sqrt 2B\sin 2\theta +2C\sin^2\theta, \nonumber\\
&&m_2\simeq A\sin^2\theta-\sqrt 2B\sin 2\theta +2C\cos^2\theta, \nonumber\\
&&m_3=0, \qquad m_4=F,
\label{eqg}
\end{eqnarray}
where we neglect the contribution from the fourth low and column of $m_\nu$ to
$m_{1,2}$ taking account of the fact such as $A~{^>_\sim}~{D^2\over F}$,
$B~{^>_\sim}~{DE\over F}$ and $C~{^>_\sim}~{E^2\over F}$.
Here we should note that in this model the violation of the proportional
relation between $(m_{2\alpha})$ and $(m_{3\alpha})$ as vectors is 
crucial to restrict a number
of zero mass eigenvalue into one and to control the mixing structure. 
There is a freedom in a choice of
two elements of $(m_{3\alpha})$ which are taken to be equal in (\ref{eqb}). 
As far as we consider the case in which the charged lepton mass 
matrix is diagonal, it is not important. But when
we consider the different situation, it might become crucial to the
consideration of the oscillation phenomena.
If the charged lepton mass matrix is diagonal, the above 
mixing matrix $U$ is just the
flavor mixing matrix $V^{\rm (MNS)}$ which controls the neutrino oscillation. 
We assume it in the charged lepton sector and also 
no $CP$ violation in the lepton sector. 
At this stage we cannot determine to which flavor each 
$\nu_\alpha $ corresponds so that we will use 
the Roman numerals for the subscript $\alpha$ for a while. 
Next we study the features of the oscillation phenomena in the model in 
order to fix the neutrino flavor.

\begin{figure}[tb]
\begin{center}
\begin{tabular}{cccc}\hline
$(\alpha,\beta)$ & $(i,j)$ & 
$-4U_{\alpha i}U_{\beta i}U_{\alpha j}U_{\beta j}(\equiv{\cal A})$  
& \\ \hline\hline
$(I,\romtw)$  & (1,2)& ${1\over 2}\sin^22\theta$  & (A)\\
$(I, \romth)$ & (1,2) &  ${1\over 2}\sin^22\theta$ & (B)\\
$(\romtw, \romth)$ & (1,3) & $\sin^2\theta$ & (C) \\
                   & (2,3) & $\cos^2\theta$ & (D) \\
                   & (1,2) & $-{1\over 4}\sin^22\theta$ & (E) \\ \hline
\end{tabular}
\end{center} 
\vspace*{1mm}

{\footnotesize Table 1.~~ The contributions to each neutrino transition
process $\nu_\alpha\rightarrow\nu_\beta$ from each sector 
$(i, j)$ of the mass eigenstates.}
\end{figure}

The transition probability due to the neutrino oscillation  
$\nu_\alpha\rightarrow \nu_\beta$ after the flight length $L$ is 
well-known to be written 
by using the matrix elements of (\ref{eqe}) as
\begin{equation}
{\cal P}_{\nu_\alpha\rightarrow \nu_\beta}(L)
=\delta_{\alpha\beta}
-4\sum_{i>j}U_{\alpha i}U_{\beta i}U_{\alpha j}U_{\beta j}
\sin^2\left({\Delta m_{ij}^2\over 4E}L\right),
\label{eqh}
\end{equation}
where $\Delta m_{ij}^2=\vert m_i^2-m_j^2\vert$ and
the weak eigenstate $\nu_\alpha$ is related to the mass eigenstate
$\tilde\nu_i$ by $\nu_\alpha=U_{\alpha i}\tilde\nu_i$
in the basis that the charged lepton mass matrix is diagonal.
In Table 1 we summalize the contribution to each neutrino transition
mode $(\alpha, \beta)$ from a sector $(i, j)$ of the 
mass eigenstates.
As a phenomenologically interesting case, we consider the 
situation that the mass 
eigenstates $\tilde\nu_1$ and $\tilde\nu_2$ are almost 
degenerate and the hierarchy
$(m_3 \ll m_1 \sim m_2) \ll m_4$ among the mass eigenvalues 
is satisfied.
This corresponds to a well-known reversed hierarchy 
scenario for the atmospheric 
and solar neutrino problems in the (3+1)-neutrino spectrum \cite{invert}. 
The absolute value of each
mass eigenvalue is smaller than the ordinarily discussed scenario 
because of $m_3=0$. Then every neutrino cannot be a hot dark matter candidate.
If we apply it to explain the atmospheric and solar neutrino data, 
the squared mass difference should be taken as \cite{oscil1,oscil2} 
\begin{eqnarray}
&&2\times 10^{-3}~{\rm eV}^2~{^<_\sim}~ 
\Delta m^2_{13}\simeq \Delta m^2_{23}~{^<_\sim}~ 6\times 10^{-3}~{\rm eV}^2, 
\label{eqha}\\
&&10^{-10}~{\rm eV}^2 ~{^<_\sim}~ \Delta m^2_{12}~{^<_\sim}~ 
1.5\times 10^{-4}~{\rm eV}^2.
\label{eqi}
\end{eqnarray} 
A suitable value of $\Delta m^2_{12}$ should be chosen within the
above range depending on which solution is adopted for the solar 
neutrino problem.

By inspecting Table 1 we find that the simultaneous explanation of
both deficits of the atmospheric neutrino and the solar neutrino is
possible if we identify the weak eigenstates of neutrinos 
$(e, \mu, \tau)$ with  $(I, \romtw, \romth)$. Under this identification the
3$\times$3 submatrix of (5) is recognized as the correctly arranged 
MNS mixing matrix.
If we note that $m_3=0$ and $\Delta m_{13}^2\simeq\Delta m_{23}^2$ are 
satisfied, we find that the atmospheric neutrino is explained by
$\nu_\mu\rightarrow\nu_\tau$ obtained as the combination of 
(C) and (D) in Table 1. 
This explanation is independent of the value of $\sin\theta$. 
On the other hand, the solar neutrino is expected to be explained by
$\nu_e\rightarrow\nu_\mu$ (A) and also 
$\nu_e\rightarrow\nu_\tau$ (B).
In both processes the amplitude ${\cal A}
(\equiv -4\sum U_{\alpha i}U_{\beta i}U_{\alpha j}U_{\beta j})$ 
is ${1\over 2}\sin^22\theta$.
Thus if $\sin^22\theta\sim 10^{-2}$, the small 
mixing angle MSW solution (SMA) is realized \cite{sma}. 
In the case of $\sin^22\theta\sim 1$, it can give the 
large mixing angle MSW solution (LMA), the low mass MSW solution (LOW)
 and the vacuum oscillation solution (VO) \cite{sma} depending on the
value of $\Delta m_{12}^2$.
The CHOOZ experiment \cite{chooz} constrains a component $U_{e3}$ 
of the MNS mixing matrix \cite{ue3}.
It comes from the fact that the amplitude ${\cal A}$ of 
the contribution to $\nu_e \rightarrow \nu_x$ with the squared 
mass differences $\Delta m_{13}^2$ 
or $\Delta m_{23}^2$ always contains $U_{e3}$. 
The model is free from this constraint since $U_{e3}=0$ is satisfied 
independently of the value of $\sin\theta$. 
\input epsf 
\begin{figure}[tb]
\begin{center}
\epsfxsize=7.6cm
\leavevmode
\epsfbox{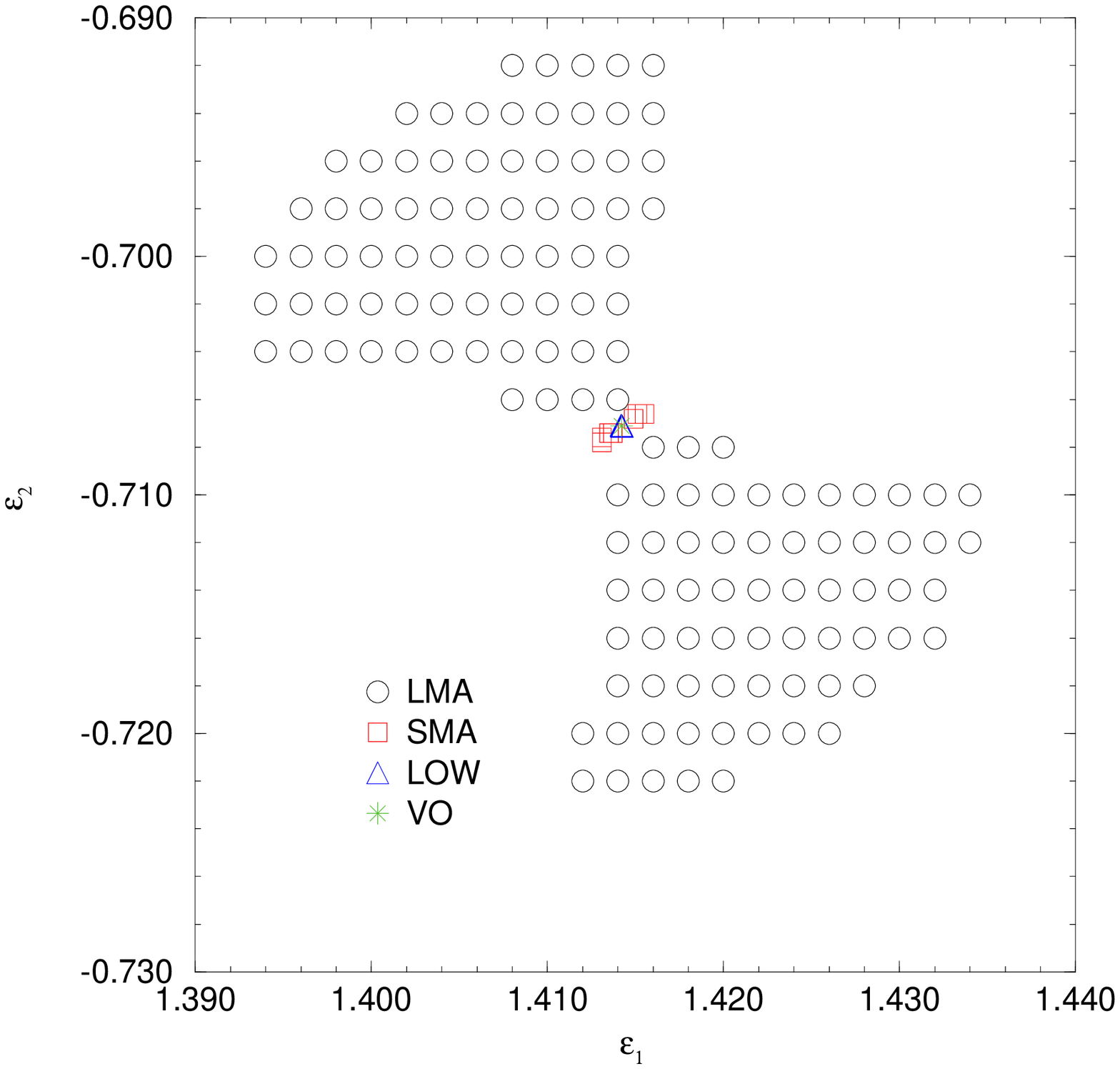}
\end{center}
\vspace*{-1.4cm}
{\footnotesize Fig. 1~~\  The scatter plot of possible solutions for
the atmospheric and solar neutrino problems in 
the $(\epsilon_1, \epsilon_2)$ plane.}
\end{figure}

In order to see the viability of the scenario in a more quantitative way 
it is useful to 
estimate numerically what kind of tuning of the primary parameters 
in (\ref{eqa}) and (\ref{eqb}) is required to realize the suitable value 
for the oscillation parameters.
For the convenience we introduce the following parametrization
for the three light states:
\begin{equation}
{\hat\eta\over \sqrt M_2}\equiv\mu^{1\over 2}, \qquad 
{\bar\eta_1\over \sqrt M_3}\equiv\epsilon_1\mu^{1\over 2}, \qquad 
{\bar\eta_2\over \sqrt M_3}\equiv\epsilon_2\mu^{1\over 2}. 
\label{eqii}
\end{equation}
For simplicity, we assume $M_2=M_3$. Then
the overall mass scale is determined by $\mu$ and the hierarchy among
the mass eigenvalues is controlled by $\epsilon_1$ and $\epsilon_2$. 
When $\epsilon_1=\sqrt 2$ and $\epsilon_2=-{1\over\sqrt 2}$, two
mass eigenvalues $m_1$ and $m_2$ are degenerate.
Using the fact, we can estimate a typical scale of $\mu$ from the condition
(\ref{eqha}) as $\mu\simeq 1.8\times 10^{-2}$~eV. This value
corresponds to $M_2\sim 5.6\times 10^{10}$~GeV for $\hat\eta\sim 1$~GeV.
The deviation from these values of $\epsilon_{1,2}$ determines the
difference between $m_1$ and $m_2$ and also the value of $\sin\theta$.
In Fig. 1 we give a scatter plot of the possible solutions for 
both the atmospheric and solar neutrino problems in the 
$(\epsilon_1,\epsilon_2)$ plane. 
In this figure solutions for both sign of $\sin 2\theta$ are contained. 
Since we consider the reversed hierarchy here, both possibilities are
allowed. From the figure we find that
the SMA, the LOW and the VO postulate the 
finer tuning of the parameters than the LMA to realize the required
values of the squared mass difference and $\sin^22\theta$.

In the present model we have a light sterile neutrino. Therefore 
we may have a chance to explain the LSND result, 
if $m_4\sim$O(1)~eV is satisfied. 
We can check whether all of mass eigenvalues and various
mixing angles quoted in the above discussion can be consistent 
with the LSND explanation by using the required
relations (\ref{eqd}), (\ref{eqf}) and (\ref{eqg}). 
If we take $\epsilon_1\sim 1.41$ and $\epsilon_2\sim -0.71$ as
a typical example in Fig. 1 and also assume $M_2=M_3$ and $m_{21}=m_{31}$, 
we obtain
\begin{equation}
\bar\eta_1\sim 1.41\hat\eta, \quad \bar\eta_2\sim -0.71\hat\eta, \quad
\sin\gamma\sim 1.21{\hat\eta\over m_{31}}, \quad
\sin\delta\sim 0.21{\hat\eta\over m_{31}}.
\label{eqkk}
\end{equation}
In order to see the feature related to the LSND we note that the 
relevant amplitude 
${\cal A}_{\rm LSND}$ can be written by using the unitarity 
of $V^{\rm (MNS)}$ and the relation 
$\vert m_{1,2,3}\vert \ll \vert m_4\vert$ as
\begin{equation}
{\cal A}_{\rm LSND}=4(V_{e4}^{\rm (MNS)})^2(V_{\mu 4}^{\rm (MNS)})^2.
\label{eqxx}
\end{equation}
Then the amplitude can be written by using eq.~(\ref{eqe}) as 
\begin{equation}
{\cal A}_{\rm LSND} \simeq 2(\cos\theta\sin\gamma
-\sin\theta\sin\delta)^2(\cos\theta\sin\delta+\sin\theta\sin\gamma)^2.
\label{eqk}
\end{equation}
The LSND data require it to be in the range around $1.2\times 10^{-3}$
for $\Delta m^2_{\rm LSND}\sim 1$~eV$^2$. 
Here we should remind that $\vert\sin\gamma\vert, ~\vert\sin\delta\vert \ll 1$ 
should be satisfied under our assumption (\ref{eqb}).
If we take the large mixing angle solutions for
the solar neutrino problem, we obtain
${\cal A}_{\rm LSND}\sim {1\over 2}(\sin^2\gamma-\sin^2\delta)^2$. 
Here we impose it to take the above mentioned value, 
we find $m_{31}\sim 5.4\hat\eta$ and then
$\sin\gamma\sim0.23$ and $\sin\delta\sim 0.04$ by using eq.~(\ref{eqkk}).
Moreover, $m_4$ can take a suitable value for the LSND result such as
$m_4\sim {2m_{31}^2\over M_2}\sim 1$~eV. 
On the other hand, if we adopt the SMA solution and then
$\cos\theta\sim 1$, we have ${\cal A}_{\rm LSND}\sim 2
\sin^2\gamma\sin^2\delta$. 
If we require it to take the suitable value, 
we find $m_{31}\sim 3.2\hat\eta$ and $m_4\sim 0.4$~eV which is too
small for the explanation of the LSND data. 
Taking account of these analyses, we find that the inclusion of the 
LSND result restricts our model to the large mixing angle solutions 
with respect to the solution for the solar neutrino problem. 
This feature of the model might be favorable if we take seriously
the recent Super-Kamiokande analysis of the solar neutrino \cite{sk}.
However, even in this case we
should comment on an influence on the big-bang nucleosynthesis (BBN) due to
the oscillation processes $\nu_{\mu,\tau}\rightarrow\nu_s$ in the early
universe. The required values of $\sin\gamma$ and $\sin\delta$ for the
explanation of the LSND data induce these processes at a large rate. 
The BBN bound on $\nu_{\mu,\tau}\rightarrow\nu_s$ given in
ref.~\cite{bbn} cannot be satisfied unless we assume the presence of
the large lepton number asymmetry at the BBN epock \cite{asym}. 

In Table. 1 an only remaining contribution (E) to 
$\nu_\mu \rightarrow \nu_\tau$
cannot imply any evidence in the short-baseline experiment even in the
case of $\sin^22\theta\simeq 1$ since $\Delta m_{12}^2$ is too small. 
However, this mode may be relevant to the long-baseline experiment 
in the case of 
$\Delta m_{12}^2\sim 10^{-4}$~eV$^2$ which corresponds to the LMA
solution of the solar neutrino deficit. We show the effect of the
mode (E) on the $P(\nu_{\mu} \rightarrow \nu_x)$ in Fig. 2.
The dashed line comes from the modes (C) and (D) which correspond
to the ordinary two flavor oscillation $\nu_\mu\rightarrow\nu_\tau$. 
The thick solid line is the one which is obtained by 
taking account of the contribution of (E). 
In the thin solid line the contribution of (A) which corresponds
to $\nu_x=\nu_e$ is also taken into account.
This shows that it may be possible to discriminate the model from 
others in the long-baseline experiment such as $L~{^>_\sim}~2000$~km. 
Moreover, the present model may be expected to have another 
experimental signature 
in the neutrinoless double $\beta$-decay \cite{bb0}.
Using eq.~(\ref{eqe}), the effective mass parameter which appears in a
formula of the rate of neutrinoless double $\beta$-decay can be estimated as
\begin{equation}
\vert m_{ee}\vert\equiv \left\vert \sum_j\vert U_{ej}
\vert^2e^{i\phi_j}m_j\right\vert
=\left(m_1\cos^2\theta+m_2\sin^2\theta\right)\sim m_1,
\label{eqj}
\end{equation}
because of the fine degeneracy between $m_1$ and $m_2$.
Thus $\vert m_{ee}\vert$ takes the value in the range of 0.04 - 0.08~eV 
which is independent of the value of $\sin\theta$, that is, the solution 
of the solar neutrino problem.
This value seems to be within the reach in the near future experiments.

\begin{figure}[tb]
\begin{center}
\epsfxsize=7.6cm
\leavevmode
\epsfbox{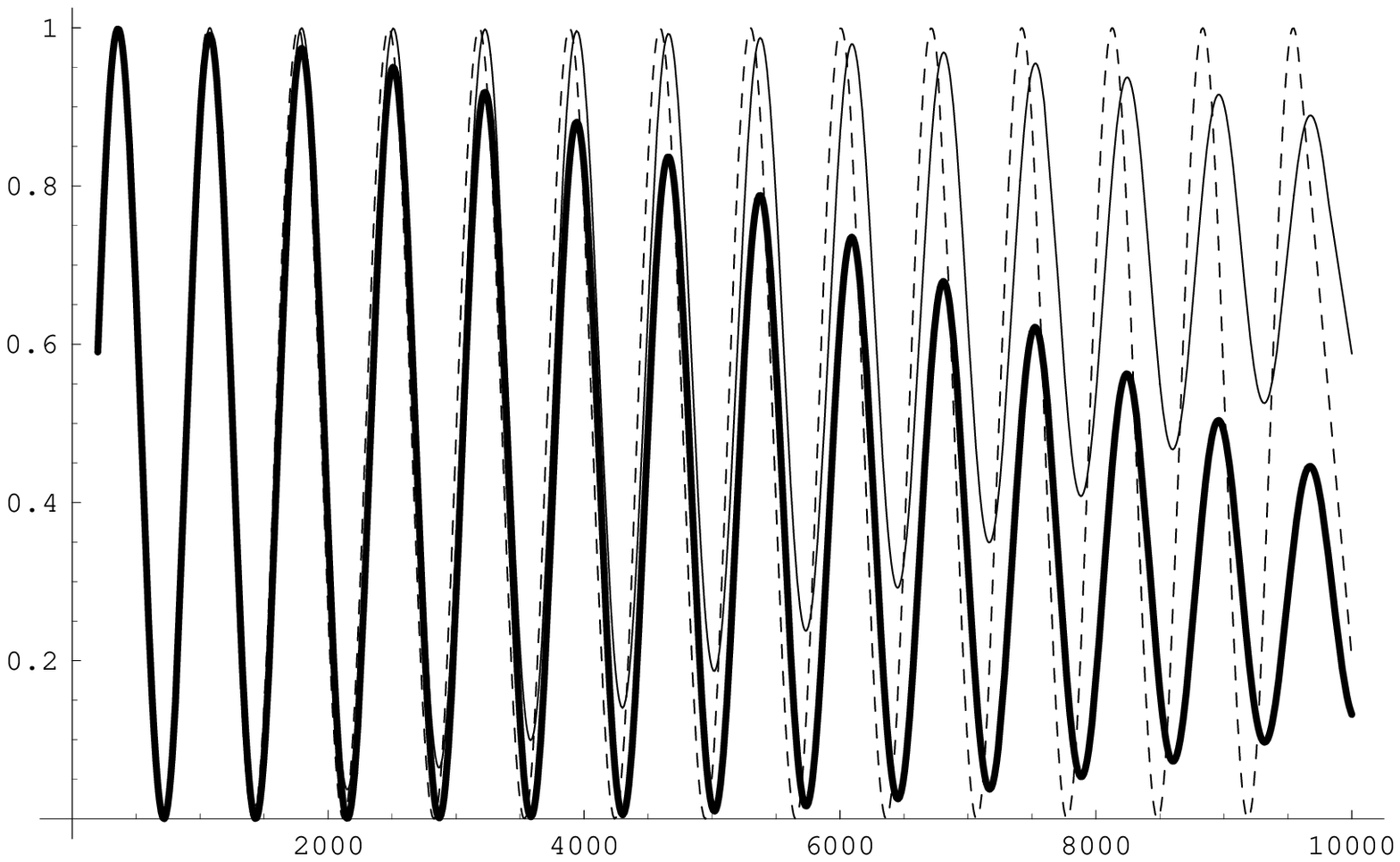}
\end{center}
\vspace*{-0.4cm}
{\footnotesize Fig. 2~~\  The transition probability
 $P(\nu_\mu\rightarrow\nu_{x(\not=\mu)})$ as a function of 
the flight length $L$~km. 
We assume $E=$1~GeV, $\Delta m_{13}^2=3.5\times 10^{-3}$~eV$^2$ and
$\Delta m_{12}^2= 10^{-4}$~eV$^2$}.
\end{figure}
Before closing this section we order some short comments.
In the above analysis we assume the reversed mass hierarchy in the (3+1) 
scheme.
As an another hierarchy among the mass eigenvalues 
we can consider the usual one, that is,
$(m_3~{^<_\sim}~ m_2 \ll m_1 ) \ll m_4$ \cite{31sch}.
However, it is found that such a hierarchy cannot be consistent with
the experimental data in the present scenario if we note that 
$\nu_e$ cannot be identified with the state $I$ to explain the solar 
neutrino deficit. 
We should also note that the above result crucially depends on the
assumption on the charged lepton sector, for which we assumed that 
its mass matrix was diagonal. As far as the flavor mixing in the
charged lepton sector is small, the above result seems to be applicable.
However, if it is large, the result can be largely changed.
When we consider the GUT, such situations happen. 
In the next section we study this issue.  
\vspace*{5mm}

\noindent
{\Large\bf 3.~Embedding into SU(5)}

We consider an embedding of our neutrino model into the supersymmetric 
SU(5) GUT. 
In that case the charged lepton mass matrix can be related to the
neutrino mass matrix through the group theoretical constraint.
Thus we cannot assume the small flavor mixing in the quark sector 
independently of the lepton sector. 
The result obtained in the previous section may be modified if 
we embed our scenario into the GUT scheme.
In order to control the flavor mixing structure we
adopt the Froggatt-Nielsen mechanism and introduce Abelian 
flavor symmetries U(1)$_{F_1}\times$U(1)$_{F_2}$.  
The symmetries are assumed to be broken by small parameters 
$\lambda$ and $\epsilon$ so that Yukawa 
couplings inducing the fermion masses are suppressed by both powers 
of $\lambda$ and $\epsilon$. As a result the fermion mass hierarchy 
is produced.
We take a model discussed in \cite{af} as such a typical
example and modify it to embed our scenario for the neutrino mass into it.

In the SU(5) GUT quarks and leptons are embedded into 
the representations of SU(5) as follows,
\begin{equation}
{\bf 10}\ni (q, u^c, e^c), \qquad {\bf 5^\ast} \ni (d^c, \ell), \qquad
{\bf 1}\ni \nu^c.
\label{eql}
\end{equation}
We assign the charge of U(1)$_{F_1}\times$U(1)$_{F_2}$ to each
representation in the following way \cite{af}:
\begin{eqnarray} 
{\bf 10}~&:&\qquad  (3, 2, 0),  \qquad (0, 0, 0), \nonumber \\
{\bf 5^\ast}~&:& \qquad (c, 0, 0), \qquad (0, 0, 0), \nonumber \\
{\bf 1}~~&:&    \qquad (0, 0, 0), \qquad (\alpha, \beta, \beta), 
\label{eqm}
\end{eqnarray}
where the numbers in the parentheses represent the charges given to each
generation and $c,\alpha$ and $\beta$ are non-negative integers.
The ordinary doublet Higgs fields $H_1$ and $H_2$ in the minimal
supersymmetric standard model are assumed to have no 
charge of U(1)$_{F_1}\times$U(1)$_{F_2}$. In addition to these fields
we introduce SU(5) singlet fields $S_1$ and $S_2$ which have the charges
$(-1, 0)$ and $(0, -1)$ of the flavor symmetries, respectively\footnote{We 
may need to introduce some fields to cancel the chiral anomaly of the
flavor symmetry if it is a non-anomalous gauge symmetry. 
However, we do not go further into this issue in the present paper.}.
The symmetries control the flavor mixing structure by regulating the
number of fields $S_1$ and $S_2$ contained in each non-renormalizable term.
If the singlet fields $S_1$ and $S_2$ get the vacuum expectation 
values $\langle S_1\rangle$ and $\langle S_2\rangle$, the above 
mentioned suppression factors for the Yukawa 
couplings can be realized as the power of
$\lambda={\langle S_1\rangle\over M_{\rm pl}}$ and $\epsilon={\langle
S_2\rangle\over M_{\rm pl}}$. Here $M_{\rm pl}$ is the Planck scale.

Using the Abelian flavor charges introduced above, we can obtain the
quark and lepton mass matrices in the following form:
\begin{eqnarray}
&&M_u\sim \left(\begin{array}{ccc}\lambda^6 &\lambda^5 &\lambda^3 \\
\lambda^5 &\lambda^4 &\lambda^2 \\ \lambda^3 &\lambda^2 & 1 \\
\end{array}\right)\langle H_2\rangle, \qquad
M_d\sim \left(\begin{array}{ccc}\lambda^{3+c} &\lambda^{2+c} & \lambda^{c} \\
\lambda^3 &\lambda^2 & 1 \\ \lambda^3 &\lambda^2 & 1 \\
\end{array}\right)\langle H_1\rangle, \nonumber \\
&&M_\nu\sim \left(\begin{array}{ccc}
\lambda^{c}\epsilon^\alpha &\epsilon^\alpha &\epsilon^\alpha \\
\lambda^{c}\epsilon^\beta &\epsilon^\beta &\epsilon^\beta \\
\lambda^{c}\epsilon^\beta &\epsilon^\beta &\epsilon^\beta 
\end{array}\right)\langle H_2\rangle, \qquad 
M_e\sim \left(\begin{array}{ccc}\lambda^{3+c} &\lambda^3 &\lambda^3 \\
\lambda^{2+c} &\lambda^2 &\lambda^2 \\ \lambda^c & 1 & 1 \\
\end{array}\right)\langle H_1\rangle, \nonumber \\
&&M_R\sim \left(\begin{array}{ccc}
\epsilon^{2\alpha} &\epsilon^{\alpha+\beta} 
&\epsilon^{\alpha+\beta} \\
\epsilon^{\alpha+\beta} &\epsilon^{2\beta} 
&\epsilon^{2\beta} \\
\epsilon^{\alpha+\beta} &\epsilon^{2\beta} 
&\epsilon^{2\beta} \\\end{array}\right)M, 
\label{eqn}
\end{eqnarray} 
where $M$ is the mass scale relevant to the origin of the right-handed
Majorana neutrino mass.  Dirac mass matrices are written in the basis 
of $\bar\psi_Rm_D\psi_L$. We do not consider the CP phases here. 
In the mass matrices (\ref{eqn}) we abbreviate the order one coupling
constants by using the similarity symbol. 
We should note that $M_\nu$ and $M_R$ in (\ref{eqn}) can have 
a similar texture to the one defined by eqs.~(\ref{eqa}) and (\ref{eqb}) 
up to the implicit coefficients of order one as far 
as $\alpha \gg\beta$ is satisfied\footnote{
In this context the order one coefficients assumed here may be allowed to be 
considered in the range $(\sqrt\epsilon,{1\over\sqrt\epsilon})$ if
$\epsilon<\lambda$ for $M_{\nu}$.}.
At least in the case of $c=0$ and 1 which is assumed 
in the following discussion, we can make $M_\nu$ satisfy the 
condition (\ref{eqb}) by tuning the order one coefficients. 
Thus we can have the similar mass matrix to eq.~(\ref{eqc}) as a result 
of the seesaw mechanism, although there is a non-zero element $M_{23}$
in $M_R$ differently from the one defined by eq.~(\ref{eqa}).
Its diagonalization matrix $U$ can be considered to have 
the similar form as eq.~(\ref{eqe}). Their difference comes only from 
the definition 
of the matrix elements $A\sim F$ as we will see it below.

In the quark sector the mass
eigenvalues and the CKM matrix elements can be found after some
inspection as
\begin{equation}
m_u : m_c : m_t = \lambda^6 : \lambda^4 : 1, \qquad
 m_d : m_s : m_b = \lambda^{3+c} : \lambda^2 : 1,
\label{eqoo}
\end{equation}
\begin{equation}
 V_{us} \sim \lambda, \qquad V_{ub}\sim \lambda^3, 
\qquad V_{cb}\sim \lambda^2. 
\label{eqo}
\end{equation}
On the charged lepton sector we can know the mass eigenvalues by 
noting the SU(5) relation such as $M_e^T=M_d$. The ratio of mass 
eigenvalues is the same as the one of the down quark 
sector and then
\begin{equation}
m_e : m_\mu : m_\tau = \lambda^{3+c} : \lambda^2 : 1.
\label{eqp} 
\end{equation}
The result has some different features from
the ones presented in ref.~\cite{af} in the down quark and charged 
lepton sectors. 
It comes from the charge assignment for ${\bf 5}^\ast$ which is
needed to realize the Dirac neutrino masses defined by
eqs. (\ref{eqa}) and (\ref{eqb}). 
If we assume $\lambda\sim 0.22$, these results seem to 
describe the experimental data in a qualitatively favorable way,
except for $m_e$ and $m_u$ which are
predicted to be too large, in particular, in the case of $c=0$.
This is the common fault known in the scheme based on the Abelian flavor 
symmetry and its similar charge assignment to the one given in (\ref{eqm}).
We cannot overcome it without something new.

We define the diagonalization matrix $\tilde U$ of the charged 
lepton mass matrix in a basis that $\tilde U^\dagger M_e^\dagger
M_e\tilde U$ is diagonal. Then $\tilde U$ can be approximately written as
\begin{eqnarray}
c=0: \quad &&\tilde U=\left(\begin{array}{cccc}
{1\over \sqrt 2}\cos\xi -{1\over \sqrt 6}\sin\xi &
{1\over \sqrt 2}\sin\xi+{1\over \sqrt 6}\cos\xi& {1\over\sqrt 3} & 0\\
-{1\over \sqrt 2}\cos\xi-{1\over \sqrt 6}\sin\xi&
-{1\over \sqrt 2}\sin\xi+{1\over \sqrt 6}\cos\xi& {1\over\sqrt 3} & 0\\
{2\over\sqrt 6}\sin\xi &-{2\over\sqrt 6}\cos\xi & {1\over\sqrt 3} & 0\\
0 & 0& 0& 1\\
\end{array}\right), \label{eqqq} \\
c=1: \quad &&\tilde U=\left(\begin{array}{cccc}
\cos\xi & 0 & \sin\xi & 0\\
-{1\over \sqrt 2}\sin\xi&
{1\over \sqrt 2}& {1\over\sqrt 2}\cos\xi & 0\\
-{1\over \sqrt 2}\sin\xi& -{1\over\sqrt 2} & 
{1\over \sqrt 2}\cos\xi& 0\\
0 & 0& 0& 1\\
\end{array}\right).
\label{eqq}
\end{eqnarray}  
The hierarchical structure (\ref{eqp}) of the mass eigenvalues requires 
a mixing angle $\xi$ to be $\sin\xi \sim O(\lambda)$.
In the neutrino sector we need to determine the finer structure 
of the Dirac neutrino mass matrix
to be suitable for the charged lepton mass matrix given in (\ref{eqn})
from a viewpoint of the explanation of various data for the neutrino 
oscillations.
For that purpose we should remind that there is a freedom in the choice of two 
elements of Dirac neutrino masses $(m_{3\alpha})$, which are taken to be equal
by tuning of the order one coefficients.
After some investigation we find that it seems to be 
favorable to take $m_{3e}=m_{3\mu }=\bar\eta_1, ~m_{3\tau }=\bar\eta_2$
instead of the one given in eq.~(\ref{eqb})\footnote{
It means that $\hat\eta\sim\bar\eta_1\sim
\lambda^c\epsilon^\beta\sim\epsilon^\beta$ and
$\bar\eta_2\sim\epsilon^\beta$. The difference among them comes from
the order one coefficients.}.
Under this assumption the mass matrix of the light neutrinos can be
written as
\begin{equation}
m_\nu=\left(\begin{array}{cccc}
A&A&B&D\\  A&A&B&D\\  B&B&C&E\\ D&D&E&F\\
\end{array}\right),
\label{eqr}
\end{equation}
and the matrix elements $A\sim F$ are defined by
\begin{eqnarray}
&&A={\hat\eta^2 \over \bar M_3}+ {\bar\eta_1^2 \over \bar M_2}
-2{\hat\eta\bar\eta_1\over \bar M_{23}}, \quad 
B={\hat\eta^2 \over \bar M_3}+ {\bar\eta_1\bar\eta_2 \over\bar M_2}
-{\hat\eta(\bar\eta_1+\bar\eta_2)\over \bar M_{23}},  \nonumber \\
&&C={\hat\eta^2 \over \bar M_3}+ {\bar\eta_2^2 \over \bar M_2}
-2{\hat\eta\bar\eta_2\over \bar M_{23}}, \quad
D={\hat\eta m_{21} \over \bar M_3}+ {\bar\eta_1 m_{31} \over \bar M_2}
-{\hat\eta m_{31}+\bar\eta_1 m_{21}\over \bar M_{23}}, \nonumber \\
&&E={\hat\eta m_{21} \over \bar M_3}+ {\bar\eta_2 m_{31} \over \bar M_2}
-{\hat\eta m_{31}+\bar\eta_2 m_{21}\over \bar M_{23}}, \quad
F={m_{21}^2 \over \bar M_3}+ { m_{31}^2 \over \bar M_2}
-2{m_{21}m_{31}\over \bar M_{23}},
\label{eqs}
\end{eqnarray}
where $\bar M_a^{-1}=M_a/(M_2M_3-M_{23}^2)$ and we partially 
use the notation in eq.~(\ref{eqb}).
The mass eigenvalues of (\ref{eqr}) are
\begin{eqnarray}
&&m_1\simeq 2A\cos^2\theta+\sqrt 2B\sin 2\theta +C\sin^2\theta, \nonumber\\
&&m_3\simeq 2A\sin^2\theta-\sqrt 2B\sin 2\theta +C\cos^2\theta, \nonumber\\
&&m_2=0, \qquad m_4=F,
\label{eqv}
\end{eqnarray}
where we again neglect the additional contributions to $m_{1,3}$ because
of the same reason as the one in the previous section.
The diagonalization matrix U is rearranged from eq.~(\ref{eqe}) because of the
change in the choice of $\bar\eta_1$ and $\bar\eta_2$. Using
the modified $U$ and eqs.~(\ref{eqqq})and (\ref{eqq}), the MNS matrix 
of the lepton mixing defined by 
$V^{\rm (MNS)}=\tilde U^TU$ is calculated for both values of $c$  as
\small
\begin{eqnarray}
&&V^{\rm (MNS)}_{c=0}\simeq\left(\begin{array}{cccc}
-{f^{(1)}_-\over \sqrt 3}\sin\xi & -\cos\xi &
{f^{(2)}_+ \over \sqrt 3}\sin\xi & 
a_1 \\
{f^{(1)}_- \over \sqrt 3}\cos\xi & -\sin\xi &
-{f^{(2)}_+ \over \sqrt 3}\cos\xi & 
a_2 \\
{f^{(2)}_+\over \sqrt 3} & 0  &
{ f^{(1)}_-\over \sqrt 3} &
a_3 \\
-\sin\gamma & 0 & -\sin\delta & 1\\
\end{array}\right), \label{eqt} \\
&&V^{\rm (MNS)}_{c=1}\simeq\left(\begin{array}{cccc}
{\cos\theta\over \sqrt 2}\cos\xi-{f^{(1)}_+ \over 2}\sin\xi & 
-{1\over\sqrt 2}\cos\xi &
-{\sin\theta\over \sqrt 2}\cos\xi-{f^{(2)}_-\over 2}\sin\xi & 
a_1 \\
{f^{(1)}_- \over 2} &{1\over 2} &
-{f^{(2)}_+\over 2} & 
a_2  \\
{\cos\theta \over\sqrt 2}\sin\xi+{f^{(1)}_+\over 2}\cos\xi & 
{1\over 2}\cos\xi &
-{\sin\theta \over\sqrt 2}\sin\xi + {f^{(2)}_- \over 2}\cos\xi& 
a_3 \\
-\sin\gamma & 0 & -\sin\delta & 1\\
\end{array}\right), \label{eqtt} 
\end{eqnarray}
\normalsize
where we use definitions
\begin{equation}
a_{i}=v_{i1}\sin\gamma+v_{i3}\sin\delta, \qquad
f^{(1)}_\pm=\cos\theta\pm\sqrt 2\sin\theta,  
\qquad f^{(2)}_\pm=\sqrt 2\cos\theta\pm\sin\theta, 
\end{equation}
and $v_{ij}$ represents the $ij$-element of the corresponding $V^{\rm (MNS)}$. 
To derive these expressions we use 
$\vert\sin\gamma\vert,~ \vert\sin\delta\vert \ll 1$ 
and neglect higher order terms of them. The mixing angles
$\theta, \gamma$ and $\delta$ in this case are defined as
\begin{equation}
\tan 2\theta ={2\sqrt 2B\over 2A-C}, \quad 
\sin\gamma\simeq {\sqrt 2D\cos\theta+ E\sin\theta\over F}, \quad
\sin\delta \simeq {-\sqrt 2D\sin\theta+E\cos\theta\over F}.
\label{equ}
\end{equation}

Now we study the oscillation phenomena in both cases in more detail.
First we consider the case of $c=0$.
Taking account that $m_2=0$ and $V_{\tau 2}^{\rm (MNS)}=0$, 
the reversed hierarchy scenario cannot be adopted from a viewpoint 
of the atmospheric neutrino problem. We must assume the normal hierarchy 
$(\vert m_2\vert ~{^<_\sim}~\vert m_1\vert \ll \vert m_3\vert )
\ll \vert m_4\vert $ 
in the (3+1)-scheme in order to realize 
$\Delta m_{12}^2\simeq \Delta m^2_{\rm solar}$ and 
$\Delta m_{23}^2\simeq \Delta m_{13}^2\simeq \Delta m^2_{\rm atm}$.
Using eq.~(\ref{eqt}), we find that the amplitude for 
$\nu_\mu\rightarrow\nu_\tau$ is
\begin{equation}
{\cal A}=-4V_{\mu 1}^{\rm (MNS)}V_{\tau 1}^{\rm (MNS)}V_{\mu 3}^{\rm (MNS)}
V_{\tau 3}^{\rm (MNS)}  
        ={4\over 9}(\sqrt 2 -{1\over 2}\tan 2\theta)^2\cos^22\theta\cos^2\xi.
\label{eqw}
\end{equation}
After some investigation we find that it suggests that $\tan 2\theta\le
0$ and $\cos^2\theta\sim 1$ should be satisfied 
for the explanation of the atmospheric neutrino problem.
We should also remind the fact that $\sin\xi\sim O(\lambda)$.
Although the large value of $\vert\sin\theta\vert$ such as 0.95 can satisfy 
the bound from the atmospheric neutrino, it seems to be disfavored 
by the solar neutrino data. Thus
the atmospheric and solar neutrino problems can be explained
by $\nu_\mu\rightarrow\nu_{\tau}$ corresponding to $\Delta m_{13}^2$ and 
$\nu_e\rightarrow\nu_{\mu}$ corresponding to $\Delta m_{12}^2$, respectively. 
The situation for the solar neutrino problem is different from 
the case in the previous section. 
Only the SMA solution is allowed in the present case
since $V_{\tau 2}^{\rm (MNS)}=0$ and $m_2=0$ make the contribution 
of $\nu_e\rightarrow\nu_\tau$ to solar neutrino deficit zero. 
It originally comes from the non-diagonal structure of the charged 
lepton mixing matrix (\ref{eqqq}). 
Since the present neutrino mass matrix induces the large mixing between 
$\nu_e$ and $\nu_\mu$ by itself,
we need the small mixing in a corresponding place of the 
charge lepton sector to realize 
the large mixing solution for the solar neutrino.
However, it is not satisfied there in this case.

In order to see the viability of the model quantitatively
we need to check numerically the consistent realization of both 
data of the atmospheric and solar neutrino observations. In this study,
for simplicity, we assume $M_{23}=0$ and $M_2=M_3$ here, although only the 
symmetries U(1)$_{F_1}\times$U(1)$_{F_2}$ cannot verify the former one. 
Then we can use the parametrization (\ref{eqii}).
In Fig.~3 we give the scatter plot of solutions for both  
the atmospheric and solar neutrino problems 
in the $(\epsilon_1, \epsilon_2)$ plane assuming 
$-0.42 \le\tan 2\theta \le 0$.
Using the figure, we can find the typical values of
the primary parameters in the model by using eqs.~(\ref{eqs}), 
(\ref{eqv}) and (\ref{equ}).
As an example, if we take $\epsilon_1\sim 3.0$ and $\epsilon_2\sim
-0.45$ from Fig.~3, we can obtain
\begin{equation}
\bar\eta_1\sim 3.0\hat\eta, \quad \bar\eta_2\sim -0.45\hat\eta, \quad
\tan 2\theta\sim -0.05 , \quad \sin\gamma \sim 2.8{\hat\eta\over
m_{31}}, \quad
\sin\delta \sim 0.35{\hat\eta\over m_{31}}.
\end{equation}
If we assume $\cos\xi=0.98$ and $\sin\xi= 0.2$, 
in this case the MNS matrix becomes
\begin{equation}
V^{\rm (MNS)}_{c=0}=\left(\begin{array}{cccc}
-0.12 & -0.98 & 0.16 & -0.12\sin\gamma+0.16\sin\delta \\
0.59 & -0.20 & -0.78  & 0.59\sin\gamma-0.78\sin\delta \\
0.80 & 0  & 0.60 &  0.80\sin\gamma+0.60\sin\delta\\
-\sin\gamma & 0 & -\sin\delta & 1\\
\end{array}\right).
\label{eqx}
\end{equation}
The CHOOZ constraint on $V_{e3}^{\rm (MNS)}$ is satisfied. 
The LSND result may have a chance to be again explained because of the
exsistence of one light sterile neutrino. In order to see it we
study the relevant amplitude ${\cal A}_{\rm LSND}$ which is estimated by 
eq.~(\ref{eqxx}).
If we require ${\cal A}_{\rm LSND}\sim 1.2\times 10^{-3}$,
we obtain $m_{31}\sim 4.7\hat\eta$ and then $m_4\sim 0.1$~eV
where we take $\mu\sim 2.6\times 10^{-3}$~eV.
This is too small to explain the LSND data. Although the larger value of 
${\cal A}_{\rm LSND}$ induces the smaller value of $m_4$,
we cannot find a favorable result for the LSND within the present freedom.  
In the present case $\mu$ and $V^{\rm (MNS)}_{e1}$ tends to take small
values by the requirement of the atmospheric and solar neutrinos. As
a result of this general feature, $m_4$ takes a small value compared to 
the required value by the LSND. 
The effective mass parameter $\vert m_{ee}\vert$ for the neutrinoless
double $\beta$-decay can be estimated as $\vert m_{ee}\vert\sim
U_{e4}^2m_4\sim 0.16\mu$. It is too small as compared with the value
expected to be reached by the near future experiment.  
\begin{figure}[tb]
\begin{center}
\epsfxsize=7.6cm
\leavevmode
\epsfbox{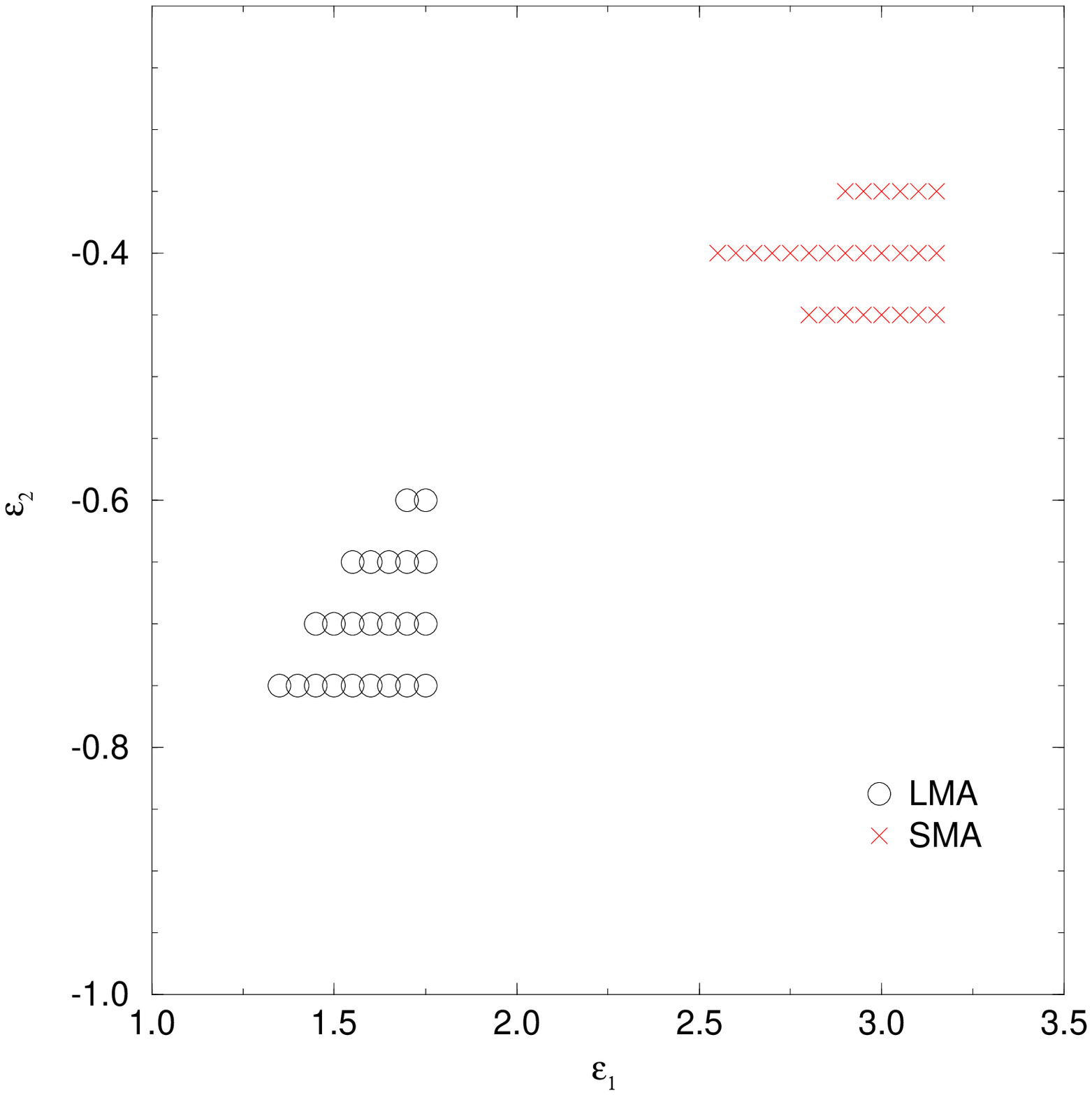}
\end{center}
\vspace*{-1.4cm}
{\footnotesize Fig. 3~~\  The scatter plot of possible solutions for
both of the solar and atmospheric neutrino problems in 
the $(\epsilon_1, \epsilon_2)$ plane. 
Requiring $-0.42\le \tan2\theta\le 0$,
the scale parameter $\mu$ is taken as $7.7\times 10^{-3}$~eV for
the LMA and $2.6\times 10^{-3}$~eV for the SMA.}
\end{figure}

It is also useful to note that the above values of the primary parameters 
of the model are realized through the suitable charge assignment 
of $\alpha$ and $\beta$.  
In order to show such an example we take $\langle H_2\rangle\sim
100$~GeV, $M\sim 4\times 10^{15}$~GeV and $\epsilon\sim 10^{-2}$. 
Then if we assign $\alpha=6$ and $\beta=0$\footnote{The more 
general condition is $\epsilon^\alpha=10^{-12}$.
Since $\epsilon$ can be taken as a very small value, it can be 
consistent even if the coefficients which are considered to
be of order one are rather large. }, we can check that 
all required quantities except for ${\cal A}_{\rm LSND}$ are 
realized in the suitable range discussed numerically above
up to the order one factors. 
In this case we have $m_4\sim 1$~eV. 

Next we treat the case of $c=1$.
Also in this case the reversed hierarchy cannot induce the sufficiently large 
amplitude ${\cal A}$ for $\nu_\mu\rightarrow\nu_\tau$ using the modes with 
$\Delta m_{12}^2$ and $\Delta m_{23}^2$. We need adopt the normal mass 
hierarchy also in the present case. The relevant amplitude for
$\nu_\mu\rightarrow\nu_\tau$ is
estimated as
\begin{equation}
{\cal A}=\sum_{i=1,2}-4V_{\mu i}^{\rm (MNS)}V_{\tau i}^{\rm (MNS)}
V_{\mu 3}^{\rm (MNS)}V_{\tau 3}^{\rm (MNS)}=
\left(\cos^2\theta-{\sin^2\theta \over 2}\right)
\left({\cos^2\theta\over 2}-\sin^2\theta+{1\over 2}\right),
\label{eqy}
\end{equation}
where we take account of $\sin\xi \ll 1$ in the estimation.
If $\cos^2\theta\sim 1$ is satisfied, the above amplitude can be
suitable to the atmospheric neutrino problem.
The contribution to the solar neutrino deficits comes from 
$\nu_e\rightarrow\nu_\mu$ and $\nu_e\rightarrow\nu_\tau$ with $\Delta
m_{12}^2$. Their combined amplitude is almost equal to one 
and the large mixing angle 
solution is realized. The value of $\cos^2\theta$ is fixed to
generate the large mixing for the explanation of the atmospheric neutrino
and it also results in the large mixing between $\nu_e$ and $\nu_\mu$.
On the other hand, in the charged lepton sector $c=1$ makes the mixing 
between $e$ and $\mu$ small so that we can have a large mixing angle solution
for the solar neutrino. 
Also in this case we give the scatter plot of the LMA solutions in Fig.~3 by
assuming $M_{23}=0$ and $M_2=M_3$ and using the parametrization (\ref{eqii}).
The LOW and VO solutions seem to be difficult to be realized since
the required $\Delta m_{12}^2$ should be much smaller than the LMA.
Using Fig.~3 we can again find the typical values of
the primary parameters in the model.
As an example, if we take $\epsilon_1\sim 1.75$ and $\epsilon_2\sim -0.75$,
we can obtain
\begin{equation}
\bar\eta_1\sim 1.75\hat\eta, \quad \bar\eta_2\sim -0.75\hat\eta, \quad
\tan 2\theta\sim -0.13 , \quad \sin\gamma \sim 1.9{\hat\eta\over
m_{31}}, \quad
\sin\delta \sim 0.26{\hat\eta\over m_{31}}.
\end{equation}
These fix the MNS matrix in the present case as
\begin{equation}
V^{\rm (MNS)}_{c=1}=\left(\begin{array}{cccc}
0.62 & -0.79 & -0.10 & 0.62\sin\gamma-0.10\sin\delta) \\
0.55 & 0.50 & -0.67  & 0.55\sin\gamma-0.67\sin\delta \\
0.58 & 0.35  & 0.73 & 0.58\sin\gamma+0.73\sin\delta\\
-\sin\gamma & 0 & -\sin\delta & 1\\
\end{array}\right).
\label{eqz}
\end{equation}
We can see that the CHOOZ constraint on
$V_{e3}^{(\rm MNS)}$ is satisfied in (\ref{eqz}).
In order to see the possibility to explain the LSND result  
we impose ${\cal A}_{\rm LSND}\sim 1.2\times 10^{-3}$.
Then by using eq.~(\ref{eqxx}) and $\mu\sim 7.7\times 10^{-3}$~eV, 
we obtain $m_{31}\sim 7.8\hat\eta$ 
and then $m_4\sim 0.93$~eV which seems to be in the suitable 
region for the LSND. 
We also have $\sin\gamma\sim 0.25$ and $\sin\delta\sim 0.03$ which
are consistent with our assumption for $\sin\gamma$ and $\sin\delta$.
These analyses show that this case can explain all neutrino
oscillation data including the LSND.
It is also interesting that the effective mass $\vert m_{ee}\vert$ for
the neutrinoless double $\beta$-decay can have rather large value
because of the $m_4$ contribution. In fact, using the above numerical
values it can be estimated as $\vert m_{ee}\vert\sim\vert
U_{e4}\vert^2m_4\sim 0.02$~eV, which may be a promising value from
the experimental viewpoint.
The value of the charges $\alpha$ and $\beta$ adopted for the $c=0$
case is also applicable to the present case to realize the model
parameters in a favored region up to the order one coefficients 
as far as we take the same values 
of $\langle H_2\rangle\sim 100$~GeV, $M\sim 10^{15}$~GeV and 
$\epsilon\sim 10^{-2}$. However, the present case needs more complicated 
structure of the order one coefficients as compared with the $c=0$ case.

Finally,  we should note that in our scheme the required equality 
among $(m_{2\alpha})$ and also among $(m_{3\alpha})$ in eq.~(\ref{eqb}) 
might not be necessary to be satisfied exactly.  
The allowed deviation should be quantitatively investigated
since it is related to the estimation of the magnitude of order 
one coefficients. 
\vspace*{5mm}

\noindent
{\Large\bf 4.~Summary}

We have proposed the scenario for the neutrino mass and mixing
based on the seesaw mechanism in the $3(\nu_L+\nu_R)$ framework. 
By assuming the special texture for the right-handed Majorana neutrino mass 
matrix and the Dirac mass matrix we could obtain a model with four light 
neutrino states including a sterile neutrino. One active neutrino is
massless and others can have the masses which are suitable for the
explanation of the atmospheric and solar neutrino deficits,
and also the LSND result.
We studied two different cases specified by the diagonal 
charged lepton mass matrix and
the non-diagonal one which is obtained by embedding of our neutrino
mass matrix into the SU(5) GUT scheme.

In the former case the so-called reversed mass hierarchy scenario
has been adopted. Every known solution for the solar neutrino 
problem could be realized by tuning the Dirac mass matrix of neutrinos. 
It is interesting that the large mixing angle MSW solution can be most 
easily realized as compared to other solutions. Moreover, if we impose
the explanation of the LSND on it, the solution for the solar neutrino problem
is restricted to the ones with the large mixing angle.
The difference from the two flavor
oscillation could be expected to be observed in the
$\nu_\mu\rightarrow\nu_\tau$ using the long-baseline experiment
with the flight length more than 2000km.
The neutrinoless double $\beta$-decay might also be accessible if the
experimental bound is improved to the level of 
$\vert m_{ee}\vert\sim$~0.04 - 0.08~eV.

In the latter case the Froggatt-Nielsen mechanism has been applied to
control the flavor mixing. 
We introduce the Abelian flavor symmetries U(1)$_{F_1}\times$U(1)$_{F_2}$
whose factor groups are assumed to have the different breaking scale.
The non-trivial charge assignment of U(1)$_{F_1}$ is used only 
for the ${\bf 10}$ and ${\bf 5}^\ast$ fields of SU(5)
and the right-handed neutrino ${\bf 1}$ is assumed to have only the 
charge of U(1)$_{F_2}$. 
Under this setting we studied the features of the mass and the mixing 
in both quark and lepton sectors for the two types of the charge assignment.
We found that it could generate the mass eigenvalues and the 
flavor mixings for the
quark sector in a qualitatively satisfactory way.
If we give the suitable flavor charge to the right-handed neutrinos, 
our neutrino scenario can be also embedded into the SU(5) scheme
consistently. 
Although the reversed mass hierarchy is disfavored in both charge 
assignments, the ordinary mass hierarchy presents a consistent explanation 
of all data of the known neutrino oscillation observations. 
For the solar neutrino problem only the SMA solution or the LMA 
solution is allowed in each case. However, if we impose the
explanation of the LSND result, only the LMA seems to be favored.

In this paper we assumed that the tuning of order one coefficients
could always be allowed. Although in our scenario the mild tuning of 
order one coefficients is very crucial, we cannot say anything on its
origin at the present stage.

\vspace*{4mm}

\noindent
{\Large\bf Acknowledgement}\\ 
This work is supported in part by the Grant-in-Aid for Scientific 
Research from the Ministry of Education, Science and Culture
(No.11640267).

\newpage

\end{document}